\documentclass[runningheads]{llncs}

\usepackage{amssymb}
\def\bbbn{{\rm I\!N}}
\def\Win{{\it W_{in}}}
\def\Wout{{\it W_{out}}}
\def\var{{\mbox{var}}}
\def\cons{{\mbox{cons}}}

\def\root{{\mbox{root}}}
\def\da{\hspace{-1.5mm}\downarrow}
\def\ua{\hspace{-1.5mm}\uparrow}

\begin{document}
\setcounter{page}{17}
\title{Value withdrawal explanations: a theoretical tool for programming environments}
\titlerunning{Value withdrawal explanations: a theoretical tool\ldots{}}
\author{Willy Lesaint}
\authorrunning{W. Lesaint}
\institute{Laboratoire d'Informatique Fondamentale d'Orl\'{e}ans \\
           rue L\'{e}onard de Vinci -- BP 6759 --
           F-45067  Orl\'{e}ans Cedex 2 -- France \\
           \email{Willy.Lesaint@lifo.univ-orleans.fr}}

\maketitle

\addtocounter{footnote}{1}
\footnotetext{In Alexandre Tessier (Ed), proceedings of the 12th International Workshop on Logic Programming Environments (WLPE 2002), July 2002, Copenhagen, Denmark.\\Proceedings of WLPE 2002: \texttt{http://xxx.lanl.gov/html/cs/0207052} (CoRR)}

\begin{abstract}
Constraint logic programming combines declarativity and efficiency
thanks to constraint solvers implemented for specific domains.
Value withdrawal explanations have been efficiently used in several
constraints programming environments but there does not exist any
formalization of them. This paper is an attempt to fill this lack.
Furthermore, we hope that this theoretical tool could help to validate
some programming environments. A value withdrawal explanation is a
tree describing the withdrawal of a value during a domain reduction by
local consistency notions and labeling. Domain reduction is formalized
by a search tree using two kinds of operators: operators for local
consistency notions and operators for labeling. These operators are
defined by sets of rules. Proof trees are built with respect to these
rules. For each removed value, there exists such a proof tree which is
the withdrawal explanation of this value. 
\end{abstract}

%%%%%%%%%%%%%%%%%%%%%%%%%%%%%%%%%%%%%%%%%%%%%%%%%%%%%%%%%%%%%%%%%%%%%%
\section{Introduction}
\label{Sect:I}
%%%%%%%%%%%%%%%%%%%%%%%%%%%%%%%%%%%%%%%%%%%%%%%%%%%%%%%%%%%%%%%%%%%%%%

Constraint logic programming is one of the most important computing
paradigm of the last years.
It combines declarativity and efficiency thanks to constraint solvers
implemented for specific domains.
Consequently the needs in programming environments is growing.
But logic programming environments are not always sufficient
to deal with the constraint side of constraint logic programming.
Value withdrawal explanations have been efficiently used in several
constraints programming environments but there does not exist any
formalization of them. This paper is an attempt to fill this lack.
This work is supported by the french RNTL\footnote{R\'eseau National
des Technologies Logicielles} project OADymPPaC\footnote{Outils pour
l'Analyse Dynamique et la mise au Point de Programmes avec
Contraintes {\tt http://contraintes.inria.fr/OADymPPaC/}} which aim is
to provide constraint programming environments.

A value withdrawal explanation is a tree describing the withdrawal of
a value during a domain reduction.
This description is done in the framework of domain reduction of
finite domains by notions of local consistency and labeling. A first
work \cite{FerLesTes-wflp-02} dealt with explanations in the
framework of domain reduction by local consistency notions only.
A value withdrawal explanation contains the whole information about a
removal and may therefore be a useful tool for programming environments.
Indeed it allows to perform:
\begin{itemize}
  \item failure analysis: a failure explanation being a set of value
        withdrawal explanations;
  \item constraint retraction: explanations provides the values which
        have been withdrawn directly or indirectly by the constraint
        and then allow to easily repair the domains;
  \item debugging: an explanation being a kind of declarative trace of
        a value withdrawal, it can be used to find an error from a
        symptom.
\end{itemize}
The first and second item have been implemented in the \texttt{PaLM}
system \cite{BarJus-trics-00}.
\texttt{PaLM} is based on the constraint solver \textsc{choco}
\cite{Lab-trics-00} where labeling is replaced by the use of
explanations.
Note that the constraint retraction algorithm of \texttt{PaLM} has
been proved correct thanks to our definition of explanations, and more
generally a large family of constraint retraction algorithms are also
included in this framework.

The main motivation of this work is not only to provide a common model
for the partners of the OADymPPaC project but also to use explanations
for the debugging of constraints programs. 
Nevertheless, the aim of this paper is not to describe the
applications of value withdrawal explanations but to formally define
this notion of explanation.

The definition of a Constraint Satisfaction Problem is given in the
preliminary section.
In third and fourth sections a theoretical framework for the
computation of solutions is described in sections \ref{Sect:DR} and
\ref{Sect:ADR}.
A computation is viewed as a search tree where each branch is an
iteration of operators.
Finally, explanations are presented in the last section thanks to the
definition of rules associated to these operators.

%%%%%%%%%%%%%%%%%%%%%%%%%%%%%%%%%%%%%%%%%%%%%%%%%%%%%%%%%%%%%%%%%%%%%%
\section{Preliminaries}
\label{Sect:P}
%%%%%%%%%%%%%%%%%%%%%%%%%%%%%%%%%%%%%%%%%%%%%%%%%%%%%%%%%%%%%%%%%%%%%%

Following \cite{Tsang-book-93}, a \emph{Constraint Satisfaction
Problem} (CSP) is made of two parts: a syntactic part and a semantic
part.
The syntactic part is a finite set $V$ of variables, a finite set $C$
of constraints and a function $\var : C \rightarrow {\cal P}(V)$,
which associates a set related variables to each constraint.
Indeed, a constraint may involve only a subset of $V$.

For the semantic part, we need to introduce some preliminary
concepts. We consider various \emph{families} $f=(f_i)_{i \in I}$.
Such a family is referred to by the \emph{function} $i \mapsto f_i$ or
by the \emph{set} $\{(i,f_i) \mid i \in I \}$.

Each variable is associated to a set of possible values.
Therefore, we consider a family $(D_x)_{x \in V}$ where each $D_x$ is
a \emph{finite non empty set}.

We define the \emph{domain} by
$\mathbb{D} = \bigcup_{x \in V} (\{x\} \times D_x)$.
This domain allows simple and uniform definitions of (local
consistency) operators on a power-set.
For domain reduction, we consider subsets $d$ of $\mathbb{D}$.
Such a subset is called an \emph{environment}.
We denote by $d|_W$ the restriction of a set $d \subseteq \mathbb{D}$
to a set of variables $W \subseteq V$, that is,
$d|_W = \{ (x,e) \in d \mid x \in W \}$.
Any $d \subseteq \mathbb{D}$ is actually a family $(d_x)_{x \in V}$
with $d_x \subseteq D_x$: for $x \in V$, we define
$d_x=\{e \in D_x \mid (x,e) \in d\}$ and call it the
\emph{environment of $x$}.

Constraints are defined by their set of allowed tuples.
A \emph{tuple} $t$ on $W \subseteq V$ is a particular environment such
that each variable of $W$ appears only once:
$t \subseteq \mathbb{D}|_W$ and
$\forall x \in W, \exists e \in D_x, t|_{\{x\}}=\{(x,e)\}$.
For each $c \in C$, $T_c$ is a set of tuples on $\var(c)$, called the
solutions of $c$.

We can now formally define a CSP.

\begin{definition}
  A \emph{Constraint Satisfaction Problem} (CSP) is defined by:
  \begin{itemize}
    \item a finite set $V$ of variables;
    \item a finite set $C$ of constraints;
    \item a function $\var : C \rightarrow {\cal P}(V)$;
    \item the family $(D_x)_{x \in V}$ (the domains);
    \item a family $(T_c)_{c \in C}$ (the constraints semantics).
  \end{itemize}
\end{definition}

Note that a tuple $t \in T_c$ is equivalent to the family
$(e_x)_{x \in \var(c)}$ and that $t$ is identified to
$\{(x,e_x) \mid x \in \var(c)\}$.

A user is interested in particular tuples (on $V$) which associate a
value to each variable, such that all the constraints are satisfied.

\begin{definition}
  A tuple $t$ on $V$ is a \emph{solution} of the CSP if
  $\forall c \in C, t|_{\var(c)} \in T_c$.
\end{definition}

\begin{example}{Conference problem}

Mike, Peter and Alan wants to give a talk on their work to each other
during three half-days. Peter knows Alan's work and vice versa. There
are four talks (and so four variables): Mike to Peter (MP), Peter to
Mike (PM), Mike to Alan (MA) and Alan to Mike (AM). Note that Mike can
not listen to Alan and Peter simultaneously (AM $\neq$ PM). Mike wants
to know the works of Peter and Alan before talking (MA $>$ AM, MA $>$
PM, MP $>$ AM, MP $>$ PM). 

This can be written in \textsc{GNU-Prolog} \cite{DiaCod-acm-00} (with
a labeling on PM) by:
\begin{verbatim}
conf(AM,MP,PM,MA):-
        fd_domain([MP,PM,MA,AM],1,3),
        MA #> AM,
        MA #> PM,
        MP #> AM,
        MP #> PM,
        AM #\= PM,
        fd_labeling(PM).
\end{verbatim}
The values $1,2,3$ corresponds to the first, second and third
half-days. Note that the labeling on $PM$ is sufficient to obtain the
solutions. Without this labeling, the solver provides reduced domains
only (no solution).

This example will be continued all along the paper.\hfill$\Box$
\end{example}

The aim of a solver is to provide one (or more) solution. In order to
obtain them, two methods are interleaved: domain reduction thanks to
local consistency notions and labeling.
The first one is correct with respect to the solutions, that is it
only removes values which cannot belong to any solution, whereas the
second one is used to restrict the search space.

Note that to do a labeling amounts to cut a problem in several
sub-problems.

In the next section, we do not consider the whole labeling (that is
the passage from a problem to a set of sub-problems) but only the
passage from a problem to one of its sub-problems. The whole labeling
will be consider in section~\ref{Sect:ADR} with the well-known notion
of search tree.

%%%%%%%%%%%%%%%%%%%%%%%%%%%%%%%%%%%%%%%%%%%%%%%%%%%%%%%%%%%%%%%%%%%%%%
\section{Domain reduction mechanism}
\label{Sect:DR}
%%%%%%%%%%%%%%%%%%%%%%%%%%%%%%%%%%%%%%%%%%%%%%%%%%%%%%%%%%%%%%%%%%%%%%

In practice, operators are associated with the constraints and are
applied with respect to a propagation queue. This method is
interleaved with some restriction (due to labeling).
In this section, this computation of a reduced environment is
formalized thanks to a chaotic iteration of operators.
The reduction operators can be of two types: operators associated with
a constraint and a notion of local consistency, and operators
associated with a restriction.
The resulting environment is described in terms of closure ensuring
confluence.

Domain reduction with respect to notions of consistency can be
expressed in terms of operators.
Such an operator computes a set of consistent values for a set of
variables $\Wout$ according to the environments of another set of
variables $\Win$.

\begin{definition}
  A \emph{local consistency operator} of type $(\Win,\Wout)$, with
  $\Win, \Wout$ $\subseteq V$ is a monotonic function
  $f: {\cal P}(\mathbb{D}) \rightarrow {\cal P}(\mathbb{D})$ such
  that: $\forall d \subseteq \mathbb{D}$,
  \begin{itemize}
    \item $f(d)|_{V \setminus \Wout} = \mathbb{D} |_{V \setminus \Wout}$,
    \item $f(d)=f(d|_{\Win})$
  \end{itemize}
\end{definition}

Note that the first item ensures that the operator is only concerned
by the variables $\Wout$. The second one ensures that this result only
depends on the variable $\Win$.

These operators are associated with constraints of the CSP.
So each operator must not remove solutions of its associated
constraint (and of course of the CSP).
These notions of correction are detailed
in~\cite{FerLesTes-RR2001-05}.

\begin{example}
In \textsc{GNU-Prolog}, two local consistency operators are associated
to the constraint \verb+MA #> PM+: the operator which reduce the
domain of MA with respect to PM and the one which reduce the
domain of PM with respect to MA.\hfill$\Box$
\end{example}

From now on, we denote by $L$ a set of local consistency operators
(the set of local consistency operators associated with the
constraints of the CSP).

Domain reduction by notions of consistency alone is not always
sufficient.
The resulting environment is an approximation of the solutions (that
is all the solutions are included in this environment).
This environment must be restricted (for example by the choice of a
value for a variable). 
Of course, such a restriction (formalized by the application of a
restriction operator) does not have the properties of correctness of a
local consistency operator: the application of such an operator may
remove solutions.
But, in the next section, these operators will be considered as a set
(corresponding to the whole labeling on a variable).
Intuitively, if we consider a labeling search tree, this section only
deals with only one branch of this tree.

In the same way local consistency operators have been defined,
restriction operators are now introduced.

\begin{definition}
  A \emph{restriction operator} on $x \in V$ is a constant function
  $f: {\cal P}(\mathbb{D}) \rightarrow {\cal P}(\mathbb{D})$ such
  that: $\forall d \subseteq \mathbb{D}, f(d)|_{V \setminus \{x\}} =
  \mathbb{D} |_{V \setminus \{x\}}$.
\end{definition}

\begin{example}
The function $f$ such that $\forall d \in \mathbb{D}, f(d) =
\mathbb{D}|_{V \setminus \{PM\}} \cup \{ ($PM$, 1) \}$ is a restriction
operator.\hfill$\Box$
\end{example}

From now on we denote by $R$ a set of restriction operators.

These two kind of operators are successively applied to the
environment.
The environment is replaced by its intersection with the result of the
application of the operator.
We denote by $F$ the set of operators $L \cup R$.

\begin{definition}
  The \emph{reduction operator} associated with the operator $f \in F$
  is the monotonic and contracting function $d \mapsto d \cap f(d)$.
\end{definition}

A common fix-point of the reduction operators associated with $F$
starting from an environment $d$ is an environment $d' \subseteq d$
such that $\forall f \in F, d'=d' \cap f(d')$, that is
$\forall f \in F, d' \subseteq f(d')$.
The greatest common fix-point is this greatest environment $d$.
To be more precise:

\begin{definition}
  The \emph{downward closure} of $d$ by $F$ is
  $max \{d' \subseteq \mathbb{D} \mid d' \subseteq d \wedge \forall f
  \in F, d' \subseteq f(d') \}$ and is denoted by
  $CL \downarrow (d,F)$.
\end{definition}

Note that $CL \downarrow (d, \emptyset) = d$ and
$CL \downarrow (d,F) \subseteq CL \downarrow (d,F')$ if
$F' \subseteq F$.

In practice, the order of application of these operators is determined
by a propagation queue. It is implemented to ensures to never forget
any operator and to always reach the closure $CL \downarrow (d, F)$.
From a theoretical point of view, this closure can also be computed by
\emph{chaotic iterations} introduced for this aim in
\cite{FagFowSol-iclp-95}.
The following definition is taken from Apt \cite{Apt-tcs-99}.

\begin{definition}
  A \emph{run} is an \emph{infinite} sequence of operators of $F$, that
  is, a run associates with each $i \in \bbbn$ $(i \ge 1)$ an element
  of $F$ denoted by $f^i$. A run is \emph{fair} if each $f \in F$
  appears in it infinitely often, that is,
  $\forall f \in F, \{i \mid f = f^i \}$ is infinite.

  The \emph{iteration} of the set of operators $F$ from the
  environment $d \subseteq \mathbb{D}$ with respect to an infinite
  sequence of operators of $F$: $f^1, f^2, \dots$ is the infinite
  sequence $d^0, d^1, d^2, \dots$ inductively defined by:
  \begin{enumerate}
    \item $d^0 = d$;
    \item for each $i \in \bbbn$, $d^{i+1} = d^i \cap f^{i+1}(d^i)$.
  \end{enumerate}
  Its \emph{limit} is $\cap_{i \in \bbbn} d^i$.

  A \emph{chaotic iteration} is an iteration with respect to a
  sequence of operators of $F$ (with respect to $F$ in short) where
  each $f \in F$ appears infinitely often.
\end{definition}

Note that an iteration may start from a domain $d$ which can be
different from $\mathbb{D}$.
This is more general and convenient for a lot of applications (dynamic
aspects of constraint programming for example).

The next well-known result of confluence
\cite{CouCou-saipl-77,FagFowSol-iclp-95} ensures that any chaotic
iteration reaches the closure.
Note that, since $\subseteq$ is a well-founded ordering
(i.e. $\mathbb{D}$ is a finite set), every iteration from
$d \subseteq \mathbb{D}$ is stationary, that is,
$\exists i \in \bbbn, \forall j \ge i, d^j=d^i$.

\begin{lemma}\label{lemma:chaotic iteration}
  The limit $d^F$ of every chaotic iteration of a set of operators $F$
  from $d \subseteq \mathbb{D}$ is the \emph{downward closure} of $d$
  by $F$.
\end{lemma}

\begin{proof}
  Let $d^0, d^1, d^2, \dots$ be a chaotic iteration of $F$ from $d$
  with respect to $f^1, f^2, \ldots$
  
  [$CL \downarrow (d,F) \subseteq d^F$]
  For each $i$, $CL \downarrow (d,F) \subseteq d^i$, by induction:
  $CL \downarrow (d,F) \subseteq d^0 = d$.
  Assume $CL \downarrow (d,F) \subseteq d^i$,
  $CL \downarrow (d,F) \subseteq f^{i+1} (CL \downarrow (d,F))
  \subseteq f^{i+1}(d^i)$ by monotonicity. Thus,
  $CL \downarrow (d,F) \subseteq d^i \cap f^{i+1}(d^i) = d^{i+1}$.

  [$d^F \subseteq CL \downarrow (d,F)$]
  There exists $k \in \bbbn$ such that $d^F=d^k$ because $\subseteq$
  is a well-founded ordering. The iteration is chaotic, hence $d^k$ is
  a common fix-point of the set of operators associated with $F$, thus
  $d^k \subseteq CL \downarrow (d,F)$ (the greatest common fix-point).
\end{proof}

In order to obtain a closure, it is not necessary to have a chaotic
iteration.
Indeed, since restriction operators are constant functions, they can
be apply only once.

\begin{lemma}
  $d^{L \cup R} = CL \downarrow ( CL \downarrow (d, R), L)$
\end{lemma}

\begin{proof}
  $d^{L \cup R} = CL \downarrow (d, L \cup R)$ by
  lemma~\ref{lemma:chaotic iteration} and
  $CL \downarrow (d, L \cup R) = CL \downarrow ( CL \downarrow (d, R),
  L)$ because operators of $R$ are constant functions.
\end{proof}

As said above, we have considered in this section a computation in a
single branch of a labeling search tree.
This formalization is extended in the next section in order to take
the whole search tree into account.

%%%%%%%%%%%%%%%%%%%%%%%%%%%%%%%%%%%%%%%%%%%%%%%%%%%%%%%%%%%%%%%%%%%%%%
\section{Search tree}
\label{Sect:ADR}
%%%%%%%%%%%%%%%%%%%%%%%%%%%%%%%%%%%%%%%%%%%%%%%%%%%%%%%%%%%%%%%%%%%%%%

A labeling on a variable can be viewed as the passage from a problem
to a set of problems.
The previous section has treated the passage from this problem to one
of its sub-problems thanks to a restriction operator. In order to
consider the whole set of possible values for the labeling on a
variable, restriction operators on a same variable must be grouped
together.
The union of the environments of the variable (the variable concerned
by the labeling) of each sub-problem obtained by the application of
each of these operators must be a partition of the environment of the
variable in the initial problem.

\begin{definition}
  A set $\{ d_i \mid 1 \le i \le n\}$ is a
  \emph{partition of $d$ on $x$} if:
  \begin{itemize}
    \item $\forall i, 1 \le i \le n, ~d|_{V \setminus \{x\}} \subseteq
          d_i|_{V \setminus \{x\}}$,
    \item $d|_{\{x\}} \subseteq \cup_{1 \le i \le n} d_i|_{\{x\}}$,
    \item $\forall i,j, 1 \le i \le n, 1 \leq j \leq n, i \neq j,
          d_i|_{\{x\}} \cap d_j|_{\{x\}} = \emptyset$.
  \end{itemize}
\end{definition}

In practice, environment reductions by local consistency operators and
labeling are interleaved to be the most efficient.

A labeling on $x \in V$ can be a complete enumeration (each
environment of the partition is reduced to a singleton) or a
splitting.
Note that the partitions always verify: 
$\forall i, 1 \leq i \leq n, d_i|_{\{x\}} \neq \emptyset$.

\begin{example}
$\{\mathbb{D}|_{V \setminus \{PM\}} \cup \{($PM$,1)\}, \mathbb{D}|_{V
\setminus \{PM\}} \cup \{($PM$,2)\}, \mathbb{D}|_{V \setminus \{PM\}}
\cup \{($PM$,3)\}$ is a partition of $\mathbb{D}$. \hfill$\Box$

\end{example}

Next lemma ensures that no solution is lost during a labeling step
(each solution will remains in exactly one branch of the search tree
defined later).

\begin{lemma}
  If $t \subseteq d$ is a solution of the CSP and
  $\{ d_i \mid 1 \le i \le n\}$ is a partition of $d$ then
  $t \subseteq \cup_{1 \leq i \leq n} CL\downarrow (d_i, L)$.
\end{lemma}

\begin{proof}
  straightforward.
\end{proof}

Each node of a search tree can be characterized by a quadruple
containing the environment $d$ (which have been computed up to now),
the depth $p$ in the tree, the operator $f$ (local consistency
operator or restriction operator) connecting it with its father and
the restricted environment $e$.
The restricted environment is obtained from the initial environment
when only the restricted operators are applied.

\begin{definition}
  A \emph{search node} is a quadruple $(d, e, f, p)$ with
  $d, e \in {\cal P}(\mathbb{D})$, $f \in F \cup \{\bot\}$ and
  $p \in \bbbn$.
\end{definition}

The depth and the restricted environment allow to localize the node in
the search tree.

There exists two kinds of transition in a search tree, those caused by
a local consistency operator which ensure the passage to one only son
and the transitions caused by a labeling which ensure the passage to
some sons (as many as environments in the partition).

\begin{definition}
  A \emph{search tree} is a tree for which each node is a search step
  inductively defined by:
  \begin{itemize}
    \item $(\mathbb{D}, \mathbb{D}, \bot, 0)$ is the root of the tree,
    \item if $(d, e, op, p)$ is a non leave node then it has:
          \begin{itemize}
            \item ever one son: $(d \cap f(d), e, f, p+1)$ with
                  $f \in L$;
            \item ever $n$ sons:
                  $(d \cap f_i(d), e \cap f_i(d), f_i, p+1)$
                  with $\{f_i(d) \mid 1 \leq i \leq n \}$ a
                  partition of $d$ and $f_i \in R$.
          \end{itemize}
  \end{itemize}
\end{definition}

\begin{definition}
  A search tree is said \emph{complete} if each leaf $(d,e,f,p)$
  is such that: $d = CL\downarrow(e,L)$.
\end{definition}

This section has formally described the computation of solvers in
terms of search trees. Each branch is an iteration of operators.

%%%%%%%%%%%%%%%%%%%%%%%%%%%%%%%%%%%%%%%%%%%%%%%%%%%%%%%%%%%%%%%%%%%%%%
\section{Value withdrawal explanations}
\label{Sect:VWE}
%%%%%%%%%%%%%%%%%%%%%%%%%%%%%%%%%%%%%%%%%%%%%%%%%%%%%%%%%%%%%%%%%%%%%%

This section is devoted to value withdrawal explanations.
These explanations are defined as trees which can be extracted from a
computation.
First, rules are associated with local consistency operators,
restriction operators and the labeling process. Explanations are then
defined from a system of such rules \cite{Aczel-handbook-77}.

From now on we consider a fixed CSP and a fixed computation.
The set of local consistency operators is denoted by $L$ and the set
of restriction operators by $R$.
The labeling introduces a notion of context based on the restricted
environments of the search node.
The following notation is used:
$\Gamma \vdash h$ with $\Gamma \subseteq {\cal P}(\mathbb{D})$ and $h
\in \mathbb{D}$. $\Gamma$ is named a \emph{context}.

Intuitively, $\Gamma \vdash h$ means $\forall e \in \Gamma, h \not\in
CL\downarrow (e,L \cup R)$. A $\Gamma$ is an union of restricted
environments, that is each $e \in \Gamma$ corresponds to a branch of
the search tree. If an element $h$ is removed in different branches of
the search tree, then a context for $h$ may contain all these
branches.

%%%%%%%%%%%%%%%%%%%%%%%%%%%%%%%%%%%%%%%%%%%%%%%%%%%%%%%%%%%%%%%%%%%%%%
\subsection{Rules}
\label{Sect:T}

The definition of explanations is based on three kinds of rules.
These rules explain the removal of a value as the consequence of other
value removals or as the consequence of a labeling.

First kind of rule is associated with a local consistency
operator. Indeed, such an operator can be defined by a system of rules
\cite{Aczel-handbook-77}. If the type of this operator is $(\Win,
\Wout)$, each rule explains the removal of a value in the environment
of $\Wout$ as the consequence of the lack of values in the environment
of $\Win$.

\begin{definition}
\label{Def:LCR}
  The set of \emph{local consistency rules} associated with $l \in L$
  is:
  \begin{center}
  \begin{tabular}{rcl}
  & $\Gamma \vdash h_1 \ldots \Gamma \vdash h_n$ & \\
  ${\cal R}_l = \{$ & $\hrulefill$ &
    $\mid \Gamma \subseteq {\cal P}(\mathbb{D}), \forall d \subseteq
    \mathbb{D}, h_1, \ldots, h_n \not\in d \Rightarrow h \not\in
    l(d)\}$ \\
  & $\Gamma \vdash h$ & \\
  \end{tabular}
\end{center}
\end{definition}

Intuitively, these rules explain the propagation mechanism. Using its
notation, the definition \ref{Def:LCR} justify the removal of $h$
by the removals of $h_1, \ldots, h_n$.

\begin{example}
$\forall e \in \mathbb{D}$, the rule
\begin{small}
\begin{center}
\begin{tabular}{ccc}
$\{e\} \vdash ($PM$,2)$ & & $\{e\} \vdash ($PM$,3)$ \\
\multicolumn{3}{c}{$\hrulefill$} \\
\multicolumn{3}{c}{$\{e\} \vdash ($AM$,1)$}\\
\end{tabular}
\end{center}
\end{small}
is associated with the local consistency operator of type
$(\{$PM$\},\{$AM$\})$ (for the constraint AM $\neq$ PM).\hfill$\Box$
\end{example}

As said above, the context is only concerned by labeling. So, here,
the rule does not modify it. Note that if we restrict ourselves to
solving by consistency techniques alone (that is without any
labeling), then the context will always be the initial environment and
can be forgotten \cite{FerLesTes-wflp-02}.

From now on, we consider ${\cal R}_L = \cup_{l \in L} {\cal R}_l$.

The second kind of rules is associated to restriction operators. In
this case the removal of a value is not the consequence of any other
removal and so these rules are facts.

\begin{definition}
  The set of \emph{restriction rules} associated with $r \in R$ is:
  \begin{center}
  \begin{tabular}{rcl}
  ${\cal R}_r = \{$ & $\hrulefill$ &
    $\mid h \not\in r(\mathbb{D}), d \subseteq r(\mathbb{D}) \}$ \\
  & $\{d\} \vdash h$ & \\
  \end{tabular}
  \end{center}
\end{definition}

These rules provide the values which are removed by a restriction.

\begin{example}
The set of restriction rules associated with the restriction operator
$r$ such that $\forall d \in \mathbb{D}, r(d) = \mathbb{D}|_{V
\setminus \{PM\}} \cup \{ ($PM$, 1) \}$ is:

\begin{small}
\begin{tabular}{cccccc}
$\{$ & $\hrulefill$ &, & $\hrulefill$  & $\}$ & with
$e_1 \subseteq r(\mathbb{D})$. \\ 
& $\{e_1\} \vdash ($PM$,2)$ & & $\{e_1\} \vdash ($PM$,3)$  \\
\end{tabular}
\end{small}\hfill$\Box$
\end{example}

This restriction ensures the computation to go in a branch of the
search tree and must be memorized because future removals may be true
only in this branch. The context is modified in order to remember that
the computation is in this branch.

From now on, we consider ${\cal R}_R = \cup_{r \in R} {\cal R}_r$.

%La derniere regle sert a collecter les informations sur une valeur
%retiree. Si durant une resolution, une valeur est retiree dans
%differentes branches de l'arbre de recherche, un arbre de preuve
%pourra être construit pour chacun de ces retraits. La regle suivante
%permet de regrouper toutes les informations concernant le retrait
%d'une valeur dans differentes branches de l'arbre de recherche dans un
%seul arbre de preuve.

The last kind of rule corresponds to the reunion of informations
coming from several branches of the search tree.

\begin{definition}
  The set of \emph{labeling rules} for $h \in \mathbb{D}$ is defined by:
  \begin{center}
  \begin{tabular}{rcl}
  & $\Gamma_1 \vdash h \ldots \Gamma_n \vdash h$ & \\
  ${\cal R}_h = \{$ & $\hrulefill$ & $\mid \Gamma_1, \ldots, \Gamma_n
    \subseteq {\cal P}(\mathbb{D}) \}$ \\
  & $\Gamma_1 \cup \ldots \cup \Gamma_n \vdash h$ & \\
  \end{tabular}
  \end{center}
\end{definition}

Intuitively, if the value $h$ has been removed in several branches,
corresponding to the contexts $\Gamma_1, \ldots, \Gamma_n$, then a
unique context can be associated to $h$: this context is the union of
these contexts.

\begin{example}
For all $e_1, e_2, e_3 \in \mathbb{D}$,
\begin{small}
\begin{tabular}{ccc}
$\{e_1\} \vdash ($MP$,2)$ & $\{e_2\} \vdash ($MP$,2)$ &
  $\{e_3\} \vdash ($MP$,2)$ \\
\multicolumn{3}{c}{$\hrulefill$} \\
\multicolumn{3}{c}{$\{e_1\} \cup \{e_2\} \cup \{e_3\} \vdash ($MP$,2)$} \\
\end{tabular}
\end{small}
is a labeling rule.
\end{example}

From now on, we consider
${\cal R}_\mathbb{D} = \cup_{h \in \mathbb{D}} {\cal R}_h$.

The system of rules ${\cal R}_L \cup {\cal R}_R \cup {\cal
R}_\mathbb{D}$ can now be used to build explanations of value
withdrawal.

%%%%%%%%%%%%%%%%%%%%%%%%%%%%%%%%%%%%%%%%%%%%%%%%%%%%%%%%%%%%%%%%%%%%%%
\subsection{Proof trees}
\label{Sect:ADP}

In this section, proof trees are described from the rules of the
previous section. It is proved that there exists such a proof tree for
each element which is removed during a computation. And finally, it is
shown how to obtain these proof trees.

\begin{definition}
  A \emph{proof tree} with respect to a set of rules
  ${\cal R}_L \cup {\cal R}_R \cup {\cal R}_\mathbb{D}$ is a finite
  tree such that, for each node labeled by $\Gamma \vdash h$, if $B$ is
  the set of labels of its children, then
  \begin{center}
  \begin{tabular}{cl}
  $B$ \\
  $\hrulefill$
  & $\in {\cal R}_L \cup {\cal R}_R \cup {\cal R}_\mathbb{D}$. \\
  $\Gamma \vdash h$ \\
  \end{tabular}
  \end{center}
\end{definition}

Next theorem ensures that there exists a proof tree for each element
which is removed during a computation.

\begin{theorem}
\label{Theoreme:ADPCL}
  $\Gamma \vdash h$ is the root of a proof tree if and only if
  $\forall e \in \Gamma, h \not \in CL\downarrow (e,R)$.
\end{theorem}

\begin{proof}
  $\Rightarrow$: inductively on each kind of rule:
  \begin{itemize}
    \item for local consistency rules, if $\forall i, 1 \le i \le n,
  h_i \not\in CL\downarrow(e_i,R)$ then $h_i \not\in CL\downarrow(e_1
  \cap \ldots \cap e_n,R)$ and so (because $h \leftarrow \{h_1,
  \ldots, h_n\} \in {\cal R}$) $h \not\in CL\downarrow(\{e_1 \cap
  \ldots \cap e_n\},R)$;
    \item for restriction rules, $h \not\in e$ so $h \not\in
  CL\downarrow (e,R)$;
    \item straightforward for labeling rules.
  \end{itemize}
  $\Leftarrow$: if $\forall i, 1 \le i \le n, h \not\in
  CL\downarrow(e_i,R)$ then (\cite{FerLesTes-RR2001-05})
  there exists a proof tree rooted by $h$ for each $e_i$. So, with
  context notion, $\forall i, 1 \le i \le n, \{e_i\} \vdash h$ is the
  root of a proof tree. Thus, thanks to the labeling rule,
  $\{e_1, \ldots, e_n\} \vdash h$ is the root of a proof tree.
\end{proof}

Last part of the section is devoted to show how to obtain these trees
from a computation, that is from a search tree.

Let us recall that $\cons(h,T)$ is the tree rooted by $h$ and with the
set of sub-trees $T$.
The traversal of the search tree is in depth first. Each branch can
then be considered separately. The descent in each branch can be
viewed as an iteration of local consistency operators and restriction
operators. During this descent, proof trees are inductively built
thanks to the rules associated to these two kind of operators
(labeling rules are not necessary for the moment).
Each node being identified by its depth, the set of trees associated
to the node $(d_p, e_p, f_p, p)$ is denoted by $S^p\da$.

These sets are inductively defined as follows:
\begin{itemize}
\item $S^0\da = \emptyset$;
\item if $f_{p+1} \in R$ then:

\vspace{-12pt}
\hspace{-11pt}
      \begin{tabular}{rcl}
      & $\{\root(t) \mid t \in T\}$ & \\
      $S^{p+1}\da = S^p\da \cup \{  \cons(\{e_p\} \vdash h, T) \mid T
        \subseteq S^p\da, h \in d_p,$ & $\hrulefill$ & $\in
        {\cal R}_{f_{p+1}} \}$ \\
      & $\{e_p\} \vdash h$ 
      \end{tabular}
\item if $f_{p+1} \in L$ then:

\hspace{-11pt}
      \begin{tabular}{rcl}
      $S^{p+1}\da = S^p\da \cup \{ \cons(\{e_{p+1}\} \vdash h,
        \emptyset) \mid h \in d_p,$ & $\hrulefill$ & $\in
        {\cal R}_{f_{p+1}} \}$ \\
      & $\{e_{p+1}\} \vdash h$
      \end{tabular}
\end{itemize}

To each node $(d, e, f, p)$ is then associated a set of proof tree
denoted by $S\da(d, e, f, p)$.

A second phase consists in climbing these sets to the root, grouping
together the trees rooted by a same element but with different
contexts. To each node $(d, e, f, p)$ is associated a new set of
proof trees $S\ua(d,e,f,p)$. This set is inductively defined:
\begin{itemize}
  \item if $(d, e, f, p)$ is a leaf then
        $S\ua(d,e,f,p) = S\da(d,e,f,p)$;
  \item if $l \in L$ then $S\ua(d, e, f, p) =
        S\da(d \cup l(d), e, l, p+1)$;
  \item if $\{ r_i(d) \mid 1 \leq i \leq n \}$ is a partition of
        $d$ then $S\ua(d,e,f,p) = S \cup S' $ with
        $S = \cup_{1 \leq i \leq n} S\ua(d \cap r_i(d), e \cap r_i(c),
        r_i, p+1)$ and

        \begin{tabular}{ccc}
        & $\{root(t) \mid t \in T\}$ & \\
        $S' = \{\cons (\Gamma \vdash h, T) \mid$ & $\hrulefill$ &
        $\in {\cal R}_\mathbb{D}, T \subseteq S \}$. \\
        & $\Gamma \vdash h$ &
        \end{tabular}
\end{itemize}

\begin{corollary}
  If the search tree rooted by $(\mathbb{D},\mathbb{D},\bot,0)$ is
  complete then  $\{\root(t) \mid t \in S \ua
  (\mathbb{D},\mathbb{D},\bot,0)\} = \{\Gamma \vdash h \mid \forall e
  \in \Gamma, h \not\in CL\downarrow(e,L)\}$.
\end{corollary}
\begin{proof}
  by theorem~\ref{Theoreme:ADPCL}.
\end{proof}

These proof trees are explanations for the removal of their root.

\begin{example}
An explanation for the withdrawal of the value $2$ from the domain of
MP can be:
\begin{small}
\begin{center}
\begin{tabular}{ccccccc}
$\hrulefill$ & & $\hrulefill$ \\
$\{e_1\} \vdash ($PM$,2)$ & & $\{e_1\} \vdash ($PM$,3)$ \\
\multicolumn{3}{c}{$\hrulefill$} & & $\hrulefill$ & & $\hrulefill$ \\
\multicolumn{3}{c}{$\{e_1\} \vdash ($AM$,1)$} &
  & $\{e_2\} \vdash ($PM$,1)$ & & $\{e_3\} \vdash ($PM$,1)$ \\
& $\hrulefill$ & & & $\hrulefill$ & & $\hrulefill$ \\
& $\{e_1\} \vdash ($MP$,2)$ & & & $\{e_2\} \vdash ($MP$,2)$ & &
  $\{e_3\} \vdash ($MP$,2)$ \\
& \multicolumn{6}{c}{$\hrulefill$} \\
& \multicolumn{6}{c}{$\{e_1\} \cup \{e_2\} \cup \{e_3\} \vdash ($MP$,2)$} \\
\end{tabular}
\end{center}
\end{small}
with $e_1$, $e_2$ and $e_3$ such that:
\begin{itemize}
\item $e_1 = \mathbb{D}|_{V \setminus \{PM\}} \cup \{($PM$,1)\}$
\item $e_2 = \mathbb{D}|_{V \setminus \{PM\}} \cup \{($PM$,2)\}$
\item $e_3 = \mathbb{D}|_{V \setminus \{PM\}} \cup \{($PM$,3)\}$
\end{itemize}
This tree must be understood as follows: the restriction of the search
space to $e_1$ eliminates the values 2 and 3 of PM. Since AM $\neq$
PM, the value 1 is removed of AM. And since MP $>$ AM, the value 2 is
removed of MP.
In the same way, the value 2 is also removed of MP with the
restriction $e_2$ and $e_3$. And finally, the root ensures that this
value is removed in each of these branches.
\hfill$\Box$
\end{example}

The size of explanations strongly depends on the consistency used, the
size of the domains and the type of constraint.
For example, if all the local consistency operators are defined from
equality constraints with arc-consistency, the size of explanations
will be minimal. On the opposite side, if all local consistency
operators are defined from inequality constraints including more than
two variables, their size will be maximal.
Note that even if the width of explanations is large, their height
remains correct in general.
It is important to recall that these explanations are a theoretical
tool. So, an implementation could be more efficient. It is possible
for example to group together values of a same variable which are
removed by the same reason.

%%%%%%%%%%%%%%%%%%%%%%%%%%%%%%%%%%%%%%%%%%%%%%%%%%%%%%%%%%%%%%%%%%%%%%
\section{Interest for Programming Environments}
\label{Sect:IPE}
%%%%%%%%%%%%%%%%%%%%%%%%%%%%%%%%%%%%%%%%%%%%%%%%%%%%%%%%%%%%%%%%%%%%%%

The understanding of solvers computation provided by the explanations
is an interesting source of information for constraint (logic)
programming environments. Moreover, explanations have already been
used in several ones. The theoretical model of value withdrawal
explanation given in the paper can therefore be an interesting tool
for constraint (logic) programming environments.

The main application using explanations concerns over-constrained
problems. In these problems, the user is interesting in information
about the failure, that is to visualize the set of constraints
responsible for this failure. He can therefore relax one of them and
may obtain a solution. 

In the PaLM system, a constraint retraction algorithm have been
implemented thanks to explanations. Indeed, for each value removed
from the environment, there exists an explanation set containing the
operators responsible for the removal. So, to retract a constraint
consists in two main steps: to re-introduce the values which contain an
operator associated to the retracted constraint in their explanation,
and to wake up all the operators which can remove a re-introduced
value, that is which are defined by a rule having such a value as
head.
The theoretical approach of the explanations have permitted to prove
the correctness of this algorithm based on explanations. There did not
exist any proof of it whereas the one we propose is
immediate. Furthermore, this approach have proved the correctness of a
large family of constraints retraction algorithms used in others
constraints environments and not only the one based on explanations.

The interest for explanations in debugging is growing. Indeed, to
debug a program is to look for something which is not correct in a
solver computation. So, the information about the computation given by
the explanations can be very precious.

They have already been used for failure analysis. In constraint
programming, a failure is characterized by an empty domain. 
A failure explanation is then a set of explanations (one explanation
for each value of the empty domain).
Note that in the PaLM system, labeling has been replaced by dynamic
backtracking based on the combination of failure explanation and
constraint retraction.

An interesting perspective seems to be the use of explanations for the
declarative debugging of constraint programs. Indeed, when a symptom
of error (a missing solution) appears after a constraint solving,
explanations can help to find the error (the constraint responsible
for the symptom). For example, if a user is expected a solution
containing the value $v$ for a variable $x$ but does not obtain any
such solution, an explanation for the removal of $(x,v)$ is a useful
structure to localize the error. The idea is to go up in the tree from
the root (the symptom) to a node (the minimal symptom) for which each
son is correct. The error is then the constraint which ensures the
passage between the node and its sons.

The theoretical model given in the paper will, I wish, bring new ideas
and solutions for the debugging in constraint programming and other
environments.

%%%%%%%%%%%%%%%%%%%%%%%%%%%%%%%%%%%%%%%%%%%%%%%%%%%%%%%%%%%%%%%%%%%%%%
\section{Conclusion}
\label{Sect:Conclusion}
%%%%%%%%%%%%%%%%%%%%%%%%%%%%%%%%%%%%%%%%%%%%%%%%%%%%%%%%%%%%%%%%%%%%%%

The paper was devoted to the definition of value withdrawal
explanations. The previous notions of explanations (theoretically
described in \cite{FerLesTes-wflp-02}) only dealt with domain
reduction by local consistency notions. Here, the notion of labeling
have been fully integrated in the model.

A solver computation is formalized by a search tree where each branch
is an iteration of operators. These operators can be local consistency
operators or restriction operators. Each operator is defined by a set
of rules describing the removal of a value as the consequence of the
removal of other values. Finally, proof trees are built thanks to
these rules. These proof trees are explanations for the removal of a
value (their root).

The interest in explanations for constraint (logic) programming
environment is undoubtedly. The theoretical model proposed here have
already validate some algorithms used in some environments and will, I
wish, bring new ideas and solutions for constraint (logic) programming
environments, in particular debugging of constraint programs.

%%%%%%%%%%%%%%%%%%%%%%%%%%%%%%%%%%%%%%%%%%%%%%%%%%%%%%%%%%%%%%%%%%%%%%
%\bibliographystyle{abbrv}
%\bibliography{/home/lesaint/recherche/biblio}

%%%%%%%%%%%%%%%%%%%%%%%%%%%%%%%%%%%%%%%%%%%%%%%%%%%%%%%%%%%%%%%%%%%%%%

\end{document}